\documentclass[aps,prl,reprint,superscriptaddress,nofootinbib]{revtex4-1}
\usepackage{graphicx}
\usepackage{amsmath}
\usepackage{amsfonts}
\usepackage{bm}
\usepackage{hyperref}
\usepackage[usenames,dvipsnames]{color}

\begin{document}
\frenchspacing

\title{Soft yet sharp interfaces in a vertex model of confluent tissue}

\author{Daniel M. Sussman}\email{dmsussma@syr.edu}
\author{J. M. Schwarz}
\author{M. Cristina Marchetti}
\author{M. Lisa Manning}
\affiliation{Department of Physics and Soft Matter Program, Syracuse University, Syracuse, New York 13244, USA}

\date{\today}

\begin{abstract}
How can dense biological tissue maintain sharp boundaries between coexisting cell populations? We explore this question within a simple vertex model for cells, focusing on the role of topology and tissue surface tension. We show that the ability of cells to independently regulate adhesivity and tension, together with neighbor-based interaction rules, lets them support strikingly unusual interfaces. In particular, we show that mechanical- and fluctuation-based measurements of the effective surface tension of a cellular aggregate yield different results, leading to mechanically soft interfaces that are nevertheless extremely sharp.
\end{abstract}

\maketitle

The process of compartmentalizing different cell populations, and maintaining those boundaries, is of vital importance in processes ranging from early embryonic development to tumor metastasis \cite{Amack2012,pawlizak2015testing}. A common paradigm, the differential adhesion hypothesis, treats each cell population as an immiscible fluid and suggests that cell sorting and compartmentalization are driven by an effective surface tension \cite{steinberg1962mechanism}, which is in turn governed by a competition between the repulsive and adhesive interactions between cells. The precise cellular mechanisms that govern effective surface tension are still under debate; some investigations suggest it is dominated by adhesive interactions~\cite{steinberg2007differential}, while others implicate actomyosin contractility~\cite{Mertz2012} or a co-regulation of these two effects~\cite{Amack2012,maitre2012adhesion,engl2014actin}. It is not even clear that different methods for measuring the effective surface tension should yield consistent results, which could explain discrepancies between observations and lead to nontrivial and unexpected dynamics for cell sorting and compartmentalization.

One hint that something interesting may be happening is a set of experiments demonstrating that many tissues can support extremely sharp boundaries between compartments or coexisting cell populations~\cite{dahmann2011boundary,landsberg2009increased,Amack2012,monier2011establishment,nnetu2012impact,Manning2010}. Here we present a possible explanation for these observations based only on the assumption that cells interact mechanically with touching neighbors, and that they might regulate these interactions differently with ``unlike'' cells. 

For simplicity, our work focuses on models for single layers of confluent cells, with no cellular gaps or overlaps. 2D vertex models represent confluent monolayers as a polygonal tiling of space where each polygon corresponds to a cell \cite{Honda1978, Ziherl2009, Idema2017}; Voronoi models, which we study here, further take the cell shapes to be given by a Voronoi tessellation of the cell positions. Vertex and Voronoi models explicitly model mechanical interactions between neighboring cells, and have successfully been used to model many biophysical processes \cite{Brodland2004, Honda2001, Manning2010, Bi2014}, ranging from embryonic development to wound healing to tumor metastasis  \cite{Hufnagel2006, Farhadifar2007, Friedl2009, Brugues2014, Etournay2015}.  We include an additional interfacial tension between different cell types to mimic the mechanical changes that are known to occur at so-called ``heterotypic'' contacts. This extra term naturally leads to a mechanism for robust cell compartmentalization. 

Surprisingly, we find that this model has a large discrepancy in different macroscopic measurements of the effective surface tension between coexisting cell populations. Specifically, we demonstrate that mechanical measurements and measurements based on the spectrum of interfacial fluctuations, which are equivalent in equilibrium particulate matter, are not equivalent in this system. This difference allows tissues to support mechanically soft interfaces that nevertheless are sharper than a fraction of a cell diameter. This result is a direct consequence of the topological nature of cellular interactions.

We begin by writing down a dimensionless form of the vertex model, a commonly used energy functional describing cells in terms of their preferred geometry,
\begin{equation}
\label{eq:energy}
e = \sum_{i=1}^{N}{\left[k_A\left(a_i - a_0\right)^2 + \left(p_i-p_0\right)^2\right]}+ \sum_{\langle i j \rangle} \delta_{[i],[j]}\gamma_{ij} l_{ij}.
\end{equation}
This energy depends on the area $a_i$ and perimeter $p_i$ of each of the $N$ cells, indexed by $i$. The unit of length is defined such that the average cell area $\langle a_i \rangle =1$. The parameter $k_A$, which we set to unity, controls the cell area stiffness relative to the perimeter stiffness. The preferred values for cell area and perimeter are $a_0$ and $p_0$, respectively. The second sum in Eq. \ref{eq:energy} introduces an explicit additional interfacial tension between unlike cells \cite{SAMOS}. The sum is over all edges, $l_{ij}$, between cells $i$ and $j$; the delta function is equal to unity if the ``type'' of cells $i$ and $j$ are different, and zero if they are the same. The parameter $p_{0}$ already represents a competition between tension and adhesion, so including explicit line tension terms allows cells to independently regulate these quantities along heterotypic interfaces. For simplicity we assume that the value of the additional tension is the same, $\gamma_{ij} = \gamma_0$, for all such interfaces. We focus on a parameter regime where the bulk is fluid-like ($p_0>3.81$). To further isolate the effect of these heterotypic interfaces, we further assume that ``unlike'' cells are identical, except for this extra tension that occurs when they touch.

There is considerable debate about the dynamical rules that best represent the motion of cells, which are fundamentally out of equilibrium. We emphasize that the unusual interfacial properties we report below \emph{do not depend} on the precise details of the equations of motion governing the system. To make this point, we carry out simulations both in and out of equilibrium. We present equilibrium data using overdamped Brownian dynamics at temperature $T$. To model out-of-equilibrium dynamics, we adopt those in self-propelled particle models \cite{Henkes2011}, where each cell tries to move along a polarization vector with self-propulsion speed $v_0$. The polarization vector rotates with a diffusion constant $D_r$ \cite{Bi2016,SAMOS,Sussman2017Rigidity}. The data we present here is restricted to a modestly out-of-equilibrium regime with $0.025 \leq v_0 \leq 0.1$ and $ 0.1 \leq D_r \leq 10$. 

Although defining an effective temperature in out-of-equilibrium systems may be problematic, we will define $T_{eff} = v_0^2/2D_r$ simply to facilitate comparisons between the dynamics schemes and display them on the same plots. In order to access reasonable system sizes and time scales, we use a GPU-accelerated simulations pack to perform the numerical work described below \cite{sussman2017cellgpu,cellGPU}.

\begin{figure}
\begin{center}
\includegraphics[width=0.4\textwidth]{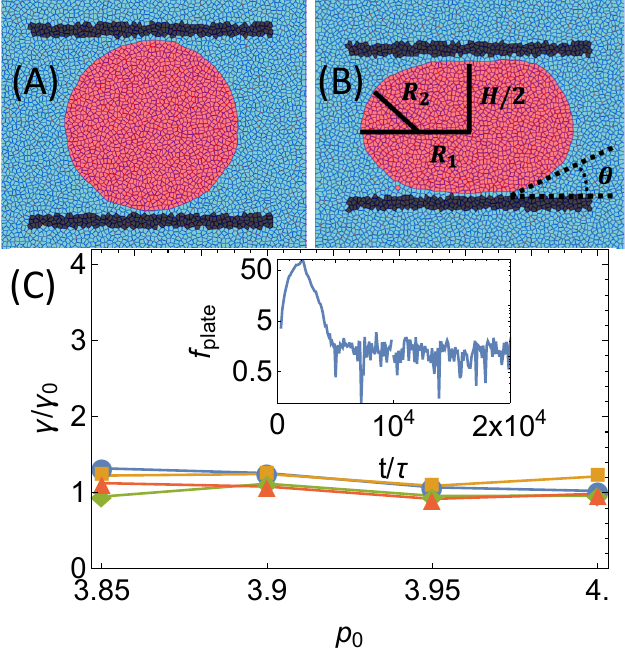}
\caption{{\bf{The effective interfacial tension measured by parallel plate compression matches the imposed microscopic tension.}} (A) Snapshot of a droplet of cells at the beginning of a parallel plate compression simulation. (B) Geometric quantities used to compute the effective surface tension, as described in the main text or via the Young-Laplace law. (C) Measured surface tension normalized by $\gamma_0$ as a function of $p_0$. Circles, squares, diamonds, and triangles correspond, respectively, to $\gamma_0 = 0.05,$  0.1, 0.2, and 0.4, with $v_0=0.1$ and $D_r=10$. This data is representative of our results for all simulations performed. [inset] Example of the mean force in the $y$-direction on lower plate, shown on a logarithmic scale, as a function of time.}
\label{fig:plateMeasurement}
\end{center}
\end{figure}

We first probe the effective surface tension of a cell droplet by numerically implementing a parallel-plate compression experiment, a popular technique in biological systems \cite{foty1994liquid,mgharbel2009measuring}. This method measures surface tensions using only the geometry of a deformed droplet and the forces exerted, i.e., without needing to measure fluid viscosities. A schematic diagram  is shown in Fig. \ref{fig:plateMeasurement}B. Although often interpreted in terms of pressure differences across the droplet interface via the Young-Laplace equation \cite{young1805essay,laplace1806traite,mgharbel2009measuring}, the force needed to maintain the plates at fixed spacing can be understood by first assuming an energy given by an effective interfacial tension, $\gamma$, times the perimeter of the droplet, and then taking the derivative of that expression. We fit the droplet shape to an ellipse with major and minor axes given by $2 R_1$ and $H$ in Fig. \ref{fig:plateMeasurement}B, and taking an analytic derivative of the perimeter of the ellipse with respect to $H$. 

We directly implement this protocol, creating parallel plates composed of externally-forced cells, with the plate-cell interactions identical to the cell-cell interactions. A sample initial configuration, showing a droplet of 1250 ``type A'' cells immersed in fluid of 3400 ``type B'' cells, is displayed in Fig. \ref{fig:plateMeasurement} A. The plates are forced together at a constant velocity over a brief time window and then held at a fixed distance apart. The total $y$-component of force on the plates is recorded; an example is shown in the inset of Fig.~\ref{fig:plateMeasurement}C. The transient behavior during the deformation protocol contains rich physical information, but to measure the effective tension we focus on the long-time limit of the external force needed to maintain the droplet in its final ellipsoidal configuration. As seen in Fig. \ref{fig:plateMeasurement}C, we find an effective surface tension consistent with the value imposed microscopically in the model. Since the additional interfacial tension we impose is smaller than the typical interfacial tensions in the bulk, we see that it leads to mechanically soft, squishy cell droplets.

We next probe the effective surface tension by measuring the spectrum of fluctuations at an interface. One common method of extracting the surface tension of a fluid phase boundary is to look at the structure and dynamics of capillary waves \cite{rowlinson2013molecular,gelfand1990finite}. In equilibrium simulations this is straightforward, as one can directly connect the interfacial roughness to the effective surface tension using standard techniques \cite{rowlinson2013molecular,Bialke2015}.

\begin{figure}
\begin{center}
\includegraphics[width=0.5\textwidth]{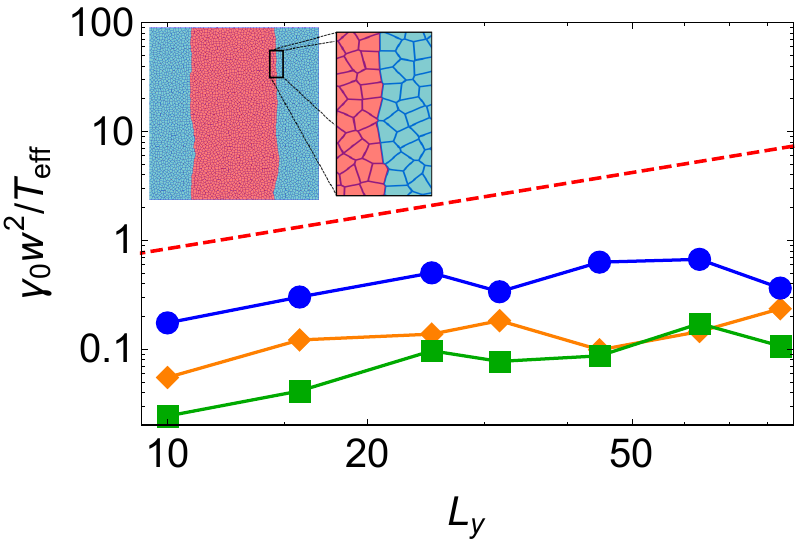}
\caption{\label{fig:strip}
{\bf{Interfaces are much sharper than expected based on the imposed microscopic tension.}} Normalized growth of the interfacial width, $w$, as a function of the size of the periodic simulation domain, $L_y$. Green squares refer to an equilibrium system at the same $T_{eff}$ as the orange diamonds, which are the self-propelled cell model with $v_0 = 0.1$, $D_r = 1$. Blue circles refer to an active system with $v_0 = 0.1$, $D_r = 0.1$. The dashed red line is the expected scaling based on the standard capillary wave argument using $\gamma=\gamma_0=0.01$ as the interfacial tension and $w_0=0$. [inset] Sample image of two cell types in a strip geometry.}
\end{center}
\end{figure}

Assuming that the Fourier spectrum of the interface has independent, Gaussian, equipartitioned modes, each height mode $h_q$ has magnitude $\langle |h_q|^2 \rangle = \frac{kT}{\gamma L_y q^2}$, where $L_y$ is the linear system size. The size of the periodic box sets the range of accessible wavevectors  $q=2\pi n/L_y$ (for $n = 1 \textrm{ to } \infty$). The mean interfacial width is  $w^2 = \sum_q \langle |h_q|^2 \rangle$, which gives 
\begin{equation} \label{eq:capillaryScaling}
w^2 = w_0^2 + \frac{kT L_y}{12\gamma},
\end{equation}
where $w_0^2$ captures the contribution of the $q=0$ mode. We simulate square domains where the length of the simulation box changes by an order of magnitude, and estimate $w$ for particular configurations by fitting the density profile of the fluid phase boundary to a hyperbolic tangent \cite{rowlinson2013molecular,Bialke2015}. We then extract the effective surface tension by evaluating the growth of the interfacial width using Eq. \ref{eq:capillaryScaling}.

In the Voronoi model this expected behavior breaks down entirely. The inset to Fig. \ref{fig:strip} shows a sample strip geometry with an interface. Visual inspection suggests that the segregation between the cell types is very sharp even for a very small imposed tension. This impression is confirmed by Fig. \ref{fig:strip}, where we plot the extracted interfacial width as a function of system size normalized by the imposed microscopic tension. For comparison, we plot a lower bound (with $w_0=0$) on the $w^2$ value expected for an equilibrium system based on our mechanical measurement, shown by the red dashed line in Fig. \ref{fig:strip}. Focusing first on the equilibrium system given by the green squares, we see the width is more than an order of magnitude smaller than expected. Equivalently, naively fitting the data would lead to an implied value of $\gamma$ nearly two orders of magnitude larger than the microscopic parameter used. Simulations with self-propelled non-equilibrium dynamics are labeled with diamonds and circles. Although these systems are still an order of magnitude sharper than expected, it seems the non-equilibrium dynamics does broaden the interface. This is consistent with observations in self-propelled particle systems \cite{Bialke2015,Speck2016}, and is an interesting avenue for future research. While we cannot rule out that longer-wavelength modes might behave more like the equilibrium expectation, the system sizes studied here (cell aggregates composed of several thousands of cells) are comparable to or larger than cell aggregates and droplets studied in experiments \cite{Manning2010}.

%
%
This reduction in the scale of interfacial fluctuations can be understood as a direct consequence of the topological nature of the interaction rules in these cell models. In particular, even though the \emph{energy} is a continuous function as cell rearrange, there are discontinuous changes in the \emph{force} whenever cells exchange neighbors. At an interface these discontinuous forces suppress fluctuations, pinning cells to the boundary and sharply compartmentalizing the cells. These forces can be analytically calculated for simplified geometries, as we show in the Supplemental Material for a square lattice of cells \cite{SupMat}.

In contrast to homogeneous vertex models where in the fluid phase four-fold vertices are generically unstable \cite{Spencer2016}, we predict that in the presence of inhomogeneous line tensions (such as those we consider in this work) some geometric configurations around four-fold vertices are stable. Although not commented upon, four-fold vertices can been seen in, e.g., the vertex-model-simulation images of Ref. \cite{landsberg2009increased}. This argument becomes slightly more complicated in the Voronoi model, as at higher values of $p_0$ the Voronoi constraints can stabilize four-fold vertices even in the bulk \cite{Sussman2017Rigidity,su2016overcrowding}. Nevertheless, we expect to see an enhancement of four-fold stabilized vertices at the interface in Voronoi models as well.

\begin{figure}
\centerline{
\includegraphics[width=0.4\textwidth]{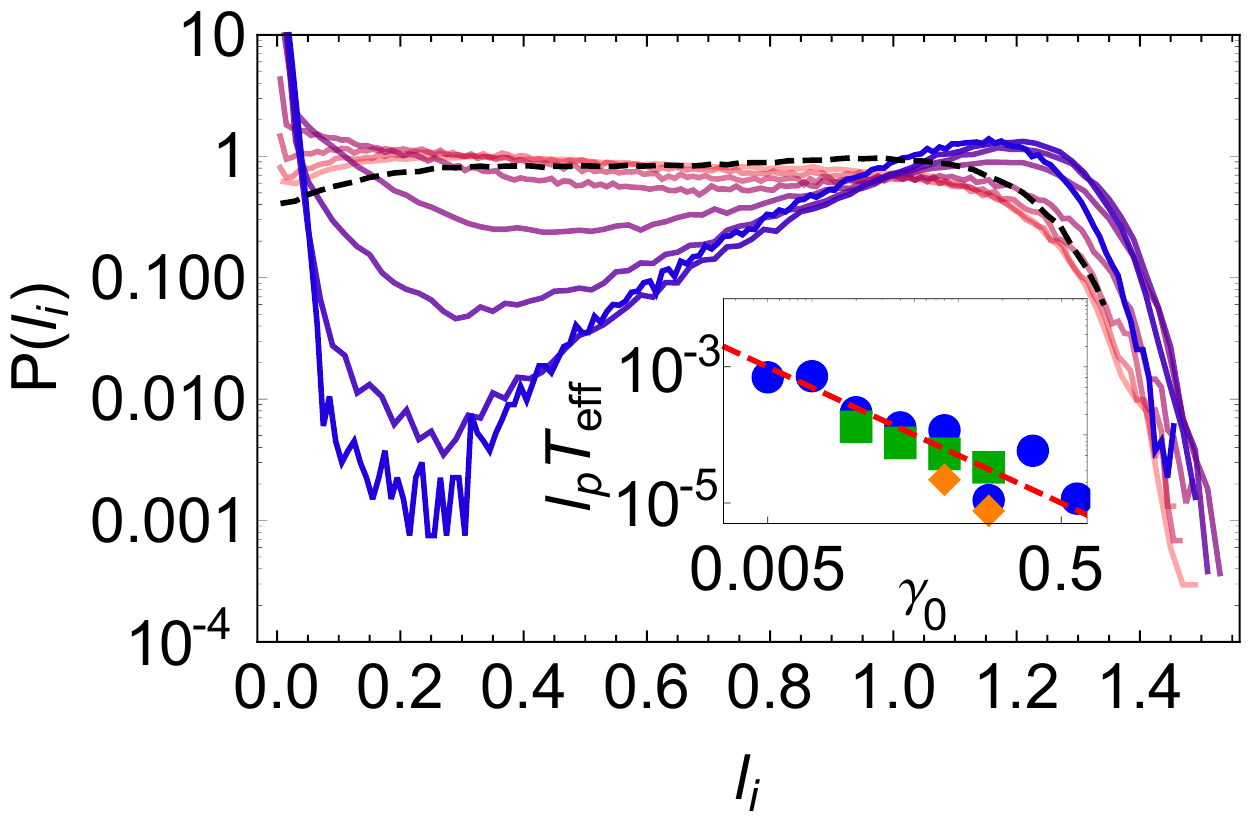}
}
\caption{\label{fig:shortEdgePlot} {\bf{Short interfacial edges become more probable with increasing tension.}} Probability distribution of interfacial edge lengths, $P(l_i)$, for $p_0=3.95$, $v_0=0.1$, $D_r = 1$ and $\gamma_0 = 0.005 - 0.64$ (light red to dark blue curves). The distribution of cell edge lengths in the bulk is given by the dashed black curve. The $l_i\ll 1$ distributions are approximately exponential, with characteristic decay length $l_p$. [inset] The length, $l_p$, characterizing the short edge distribution for $D_r = 1$ and $v_0=0.1,\ 0.05,\ 0.025$. The dashed red line is a guide to the eye with slope -1, corresponding to a Boltzmann expectation based on the energy of interfacial edges.}
\end{figure}

At finite effective temperatures these four-fold vertices will transiently resolve into three-fold vertices separated by a short edge. Therefore, we test our predictions by comparing the distribution of edge lengths $l_i$ in the bulk and along the interface between two cell types. Figure \ref{fig:shortEdgePlot} shows the bulk distribution (black dashed line) and the interfacial distribution for a wide range of imposed tensions (solid lines). For any value of applied tension we see a clear enhancement of very short interfacial edges with respect to the bulk distribution, indicating that the Voronoi tessellation is approximating an excess population of four-fold vertices. There is also an excess of slightly larger-than-normal interfacial edges, since the strip boundary must span the simulation box. As shown in the inset to Fig. \ref{fig:shortEdgePlot}, for very short edges we see that the interfacial edge length distribution is approximately exponential, with a decay length $l_p$ consistent with simple energetic considerations based on $\gamma_0$ and $T_{eff}$.

\begin{figure}
\centerline{
\includegraphics[width=0.45\textwidth]{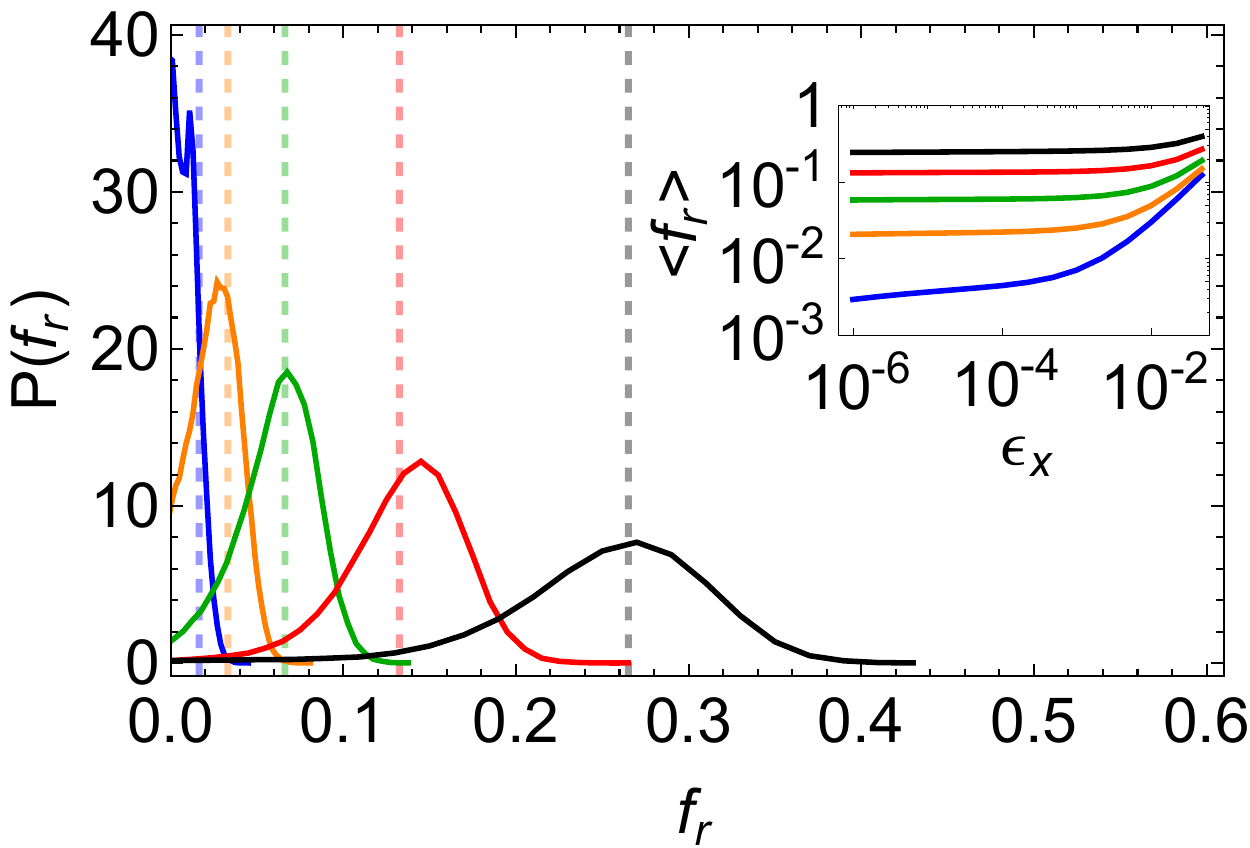}
}
\caption{\label{fig:restoringForcePlot} {\bf{The restoring force changes discontinuously when interfacial cells are displaced.}} The distribution of restoring forces, $P(f_r)$, on cells displaced by $\epsilon_x=10^{-4}$ in the inherent state of a system prepared in a strip geometry with $p_0 = 3.95$, $D_r=1$, $v_0=0.1$. From left to right $\gamma_0 = 0.04,\  0.08,\ 0.16,\ 0.32,\ 0.64$. The dashed vertical lines show the magnitude of the restoring force predicted from the analytical calculation in the Supplemental Material based on breaking four-fold vertices in a square-lattice geometry for each value of $\gamma_0$. [inset] For each value of $\gamma_0$ the mean of the restoring force is nearly independent of the displacement for small displacements, demonstrating the non-Hookean behavior of the interface.}
\end{figure}

We next show that the near-four-fold vertices in the disordered geometries in our simulations give rise to discontinuous restoring forces that suppress fluctuations of the interface. We quantify this looking at the inherent state of a system and systematically displacing each cell at the interface and measuring the restoring force, shown in the inset to Fig. \ref{fig:restoringForcePlot}, to measure the forces in the underlying energy landscape. At a particulate interface governed by adhesive interactions one would expect a spring-like restoring force proportional to the magnitude of the displacement. Instead, we find that the mean force is nearly independent of the displacement over orders of magnitude, and that it is proportional to the applied $\gamma_0$, precisely as predicted by our analysis of the simplified geometry calculated in the supplemental material. The main panel of Fig. \ref{fig:restoringForcePlot} shows the full distribution of restoring forces at a specific value of the displacement $\epsilon_x = 10^{-4}$, compared to our analytic prediction (dashed lines), indicating that our simple model captures the origin of anomalous behavior in the simulations.

We have demonstrated that adding a simple interfacial tension term in a model of topologically interacting cells can lead to highly non-trivial material behavior. We find a strong discrepancy between the effective surface tension defined by mechanical measurements versus those based on fluctuations, even in equilibrium. The roughness of interfaces is almost completely suppressed, leading to strikingly sharp boundaries between fluid domains.

How is it possible that a mechanical measurement gives a different answer than the fluctuation-based one? Our works suggests that the nature of the mechanical measurement is important: the parallel plate experiment accesses much larger forces because it is strain-controlled, and so it can overcome cusp-like pinning forces. The signature of this is likely contained in the transient behavior of the forces shown in the inset of Fig. \ref{fig:plateMeasurement}B. A small-scale microrheology experiment, for example, might yield a result more similar to the fluctuation-based measurement.

The extreme interfacial sharpening is due to the topological nature of the intercellular interactions. Since cells interact with their neighbors, and not according to the distance between cells, a cusp-like energy landscape underlies the dynamics. We have confirmed that this behavior is robust to changes to single cell mechanics (e.g., $p_0$) and changes to the propulsion mechanism (e.g., thermal vs. self-propelled). This interfacial sharpening mechanism has obvious implications for cell compartmentalization, but it also may influence the process of cell sorting. It is commonly assumed that both compartmentalizing and sorting proceed as if cells were immiscible fluids; we have seen that many-fold vertices may fundamentally alter compartmentalization, and we speculate that they may likewise have profound consequences for the process by which cells sort.

Recent work has suggested that surprising consequences can arise from systems interacting via topological interactions rather than purely metric-based ones \cite{ballerini2008interaction,ginelli2010relevance,herbert2011inferring,peshkov2012continuous}. While much of the work in this direction has focused on explicitly non-equilibrium systems -- animal flocking or self-propelled particles interacting with combined Viscek and Voronoi dynamics -- here we have shown that surprising interfacial behavior may arise as a generic consequence of the cusp-like landscape generated by the topological rules. We speculate that, in the context of real cellular aggregates, epithelial cells may interact topologically whereas mesenchymal or non-confluent cells may interact metrically (through the surrounding medium or otherwise). In addition to its relevance for confluent cellular aggregates, our findings may point towards a interesting new class of bio-inspired materials, where combining mesoscopic interaction units with independent regulation of tension and adhesion may support a diverse set of unusual material properties.

\begin{acknowledgments}
This work was primarily supported by NSF-POLS-1607416. Additional support was provided by the Simons Foundation Targeted grant 342354 (MCM) and Investigator grant 446222 (MLM) in the Mathematical Modeling of Living Systems, the National Science Foundation awards DMR-1609208 (MCM), DMR-1507938 (JMS) and DMR-1352184 (MLM), and a Cottrell Scholar award from the Research Corporation for Science Advancement (MLM). All authors acknowledge support of the Syracuse University Soft Matter Program and computing support through NSF NSF ACI-1541396. The Tesla K40 used for this research was donated by the NVIDIA Corporation.
\end{acknowledgments}

\bibliography{VoronoiInterface_bib}

\newpage
\appendix
\section{Appendix: interface stability}

\begin{figure}
\centerline{
\includegraphics[width=0.3\textwidth]{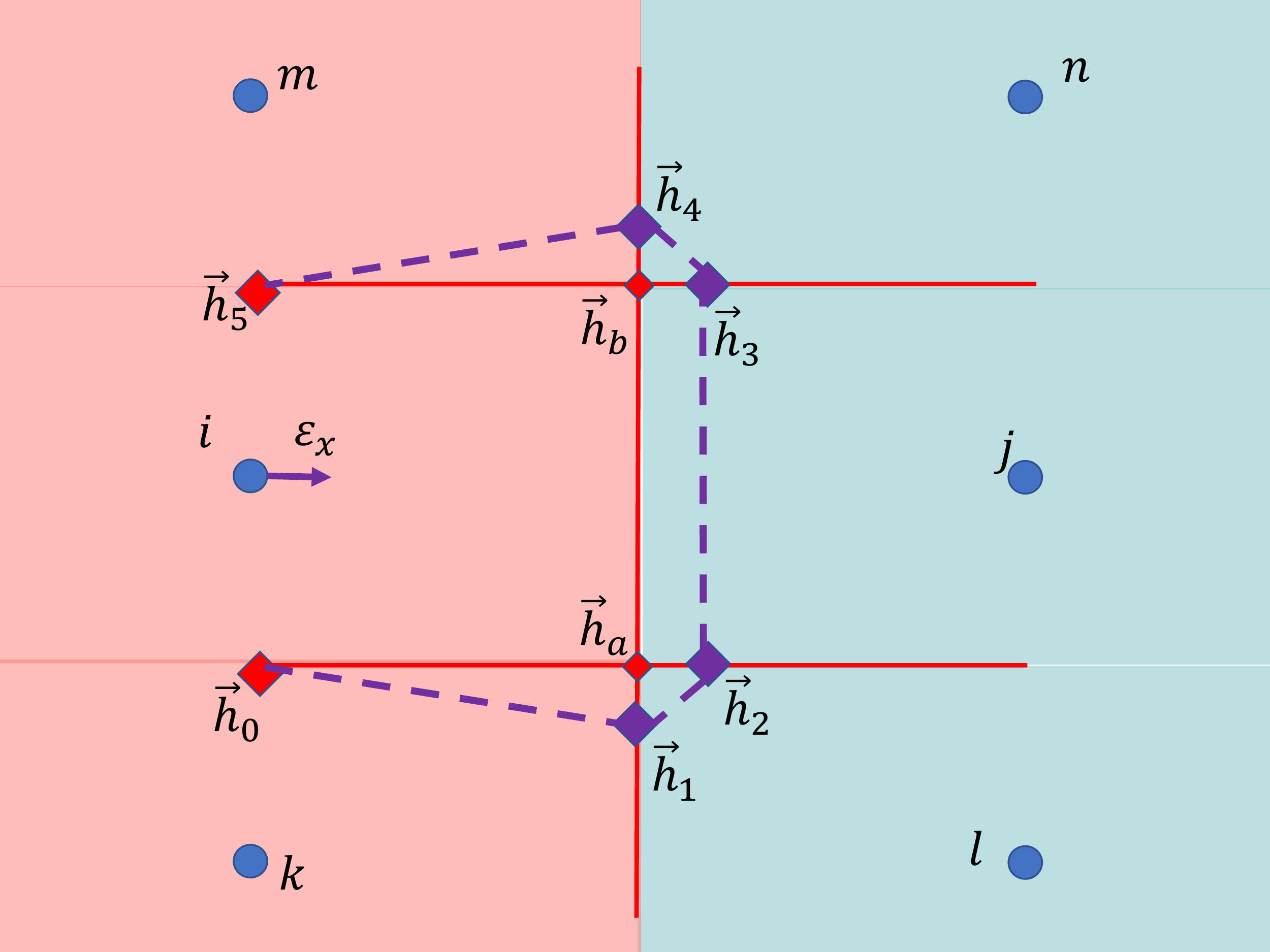}
}
\caption{\label{fig:schematic} A cartoon geometry in which six Voronoi cells (positions indicated by the blue circles) lie on a square lattice, with a vertical interface between cells of unlike type.. After displacing cell $i$ by an amount $\epsilon_x$, the new Voronoi vertices defining the cells are shown with diamond marker connected by dashed lines. In this idealized geometry an exact accounting of the initial and final energy of the drawn configuration can be analytically performed.}
\end{figure}

To more explicitly understand the interfacial sharpness between fluid-fluid mixtures in the Voronoi model with tension, we perform a straightforward calculations of the forces on cells in the square-lattice geometry shown in Fig. \ref{fig:schematic}. The six cells lie on a unit-length square lattice, cells $k,\ i,\ ,m$ on the left are of one type and cells $l,\ j,\ n$ on the right are of a different type (three additional cells bordering cell $i$ and of its type are not shown, but serve to complete the set of neighbors defining the Voronoi tessellation).

As indicated in Fig. \ref{fig:schematic}, we consider a small horizontal displacement of cell $i$ by $\epsilon_x \hat{x}$. We calculate the new Voronoi tessellation geometry, from which we can calculate the change in the energy. The initial energy of the square-lattice configuration of cell $i$ and the surrounding eight cells is
\begin{equation}
E_i = 9 k_A(a_i-a_{0})^2 + 9(p_i-p_{0})^2 + 3 \gamma_0 l_{ij},
\end{equation}
where $l_{ij} = 1$ for this geometry. We further simplify by assuming that the cells preferences are commensurate with a square lattice geometry being the energy minima, i.e. $A_0=1$ and $p_0=4$, with the line tension supported by the boundary conditions. With this simplification the initial energy is just 
\begin{equation}
E_i = 3 \gamma_0.
\end{equation}
After displacing the cell as indicated, the new total energy is, to leading order,
\begin{eqnarray}
E_f &=& 3 \gamma_0 + \gamma_0(\sqrt{2}-1)\epsilon_x \\
&&+ \left(\frac{3k_A}{2} + (22-14\sqrt{2})+\gamma_0(1-1/\sqrt{2}) \right)\epsilon_x^2\nonumber
\end{eqnarray}

In this simple calculation we see that while the energy continuously changes with cell displacement, the energy minimum is a cusp, with linear rather than quadratic growth as the lowest order term. This translates to a discontinuous jump in the restoring force,  $\vec{F} = -\nabla E$, pushing cell $i$ back towards a position supporting four-fold coordinated vertices. Thus, in the Voronoi model even a small line tension term can stabilize some number of four-fold vertices along a fluid-fluid interface. Note that in non-Voronoi vertex models it is straightforward to show that individual line tension terms between different cells can stabilize a population of multi-fold vertices \cite{Spencer2016}; this can be seen in the images of, e.g., Ref. \cite{landsberg2009increased} of vertex model simulations between cells with additional line tension terms between them. Generically, perturbing the interface, even to linear order, leads to a splitting of these four-fold interfaces, which in turn results in a discontinuous restoring force acting to restore the interfacial sharpness.

In the language of the standard capillary wave scaling argument, the energetic cost of a sinusoidal perturbation of the interface of amplitude $\epsilon$   is assumed to have an energetic cost of $\sim \epsilon^2$. With this explicit line tension term and the breaking of four-fold vertices, though, the effective energetic cost is proportional to the number of four-fold vertices along the interface, which we numerically observe to be simply proportional to the length of the interface itself (i.e. corresponding to a finite fraction of all cells along the interface), $\sim |\epsilon|$, where the constant of proportionality is determined by the material details of the system under study. Equivalently, each mode $h_q$ incurs an extensive effective energetic cost, $\propto L_y$, and thus the interfacial width is kept microscopically sharp.

\end{document}